\documentclass[fleqn,12pt,twoside]{article}
\usepackage[]{espcrc1}

\def\be{\begin{equation}} \def\ee{\end{equation}}
\def\bq{\begin{eqnarray}} \def\eq{\end{eqnarray}}
\def\p{\bullet}

\readRCS
$Id: espcrc1.tex,v 1.2 2004/02/24 11:22:11 spepping Exp $
\usepackage{graphicx}

\usepackage[figuresright]{rotating}

\newcommand{\AmS}{{\protect\the\textfont2
  A\kern-.1667em\lower.5ex\hbox{M}\kern-.125emS}}

\title{Gamma Ray Bursts and the transition to Quark Matter in Compact Stars}

\author{A. Drago,
 \address{Dipartimento di Fisica, Universit{\`a} di Ferrara and\\INFN, Sezione di Ferrara, 44100 Ferrara, Italy}
 A. Lavagno
\address{Dipartimento di Fisica, Politecnico di Torino and\\INFN,
Sezione di Torino, 10129 Torino, Italy}
G. Pagliara 
\address{Dipartimento di Fisica, Politecnico di Torino and\\INFN,
Sezione di Ferrara, 44100 Ferrara, Italy}
}

\begin{document}

\maketitle

\begin{abstract}

We discuss a model for long Gamma-Ray-Bursts in which the central 
engine is associated with the conversion process of a metastable hadronic star into a star containing 
quark matter. We analyze also the observational signatures of the model,
i.e. the Supernova-GRB temporal connection and the existence of long quiescent times in the 
temporal structure of Gamma-Ray-Bursts. 

\end{abstract}

\vspace{1cm}

Several observations indicate that long
Gamma Ray Bursts (GRBs) are connected to the final stage of
massive stars. In a few
cases a direct association between a Supernova and a GRB 
has been found but it has not yet been
clarified if the two explosions are always simultaneous or if a 
time delay can exist, with the SN preceding
the GRB. 

In one of the most popular models,
the Collapsar model \cite{woosley}, GRBs are generated by
relativistic jets from massive helium stars whose cores have collapsed
to a black hole and an accretion disk. In the Collapsar model the beaming of
GRBs is naturally explained by the ``funnel mechanism''. The crucial ingredient
of the model is a huge initial angular momentum 
of the star. One of the predictions of the Collapsar model is the SN-GRB connection 
with a short time delay ($< 100 $ s) between the two explosions.   

Here we discuss a quark deconfinement model 
\cite{noiapj,noiprd} in which 
the energy source of the GRB emission is the process of conversion
from a metastable, purely hadronic star into a more compact
star in which deconfined quark matter is present.  In our scenario,
we assume a finite value of the surface tension between hadronic and
quark matter. Therefore the hadronic star can become metastable and its mean-life time is
related to the time needed to nucleate a drop of quark matter. The
time delay between the birth of the hadronic star and the subsequent
conversion into an hybrid or quark star corresponds to the delay
between the SN explosion and the GRB. Temperature has no
effect in our scheme because we assume that when quark matter forms the
temperature is so low \cite{Pons} that only quantum tunneling is a practicable
mechanism \cite{iida}. The central density of a pure hadronic star can then
increase, due to spin down or mass accretion, until its value
approaches the deconfinement critical density. At this point a
virtual drop of quark matter can form but its nucleation time
can be extremely long. 

\begin{table}[t]
\begin{center}
\tabcolsep=0.1\tabcolsep 
\begin{tabular}{ccccccccc}
\hline
\hline
Hadronic         &
$B^{1/4}$        & 
$\sigma$         & 
$M_{cr}/M_\odot$ & 
$\Delta E$  & 
$\Delta E$  &  
$\Delta E$  & 
$\Delta E$  &
$\Delta E$  \\
Model        &
[MeV]        &
[MeV/fm$^2$] & 
             & 
$\Delta=0$   &  
$\Delta_1$   &   
$\Delta_2$   &  
$\Delta_3$   &
$\Delta_4$   \\

\hline
GM3 & $170$  & $10$ & $1.12$  & $18$  &  $52$   &  $57$   & $86$ & $178^\p$  \\
GM3 & $170$  & $20$ & $1.25$  & $30$  &  $66$   &  $72$   & $106$ & $205^\p$   \\
GM3 & $170$  & $30$ & $1.33$  & $34$  &  $75$   &  $81$   & $120$ & $221^\p$   \\
GM3 & $170$  & $40$ & $1.39$  & $38$  &  $82$   &  $88$   & $131$ & $234^\p$   \\
\hline
GM3 & $180$  & $10$ & $1.47$  & BH  &  $35$   &  $38$   & BH & --  \\
GM3 & $180$  & $20$ & $1.50$  & BH  &  $38$   &  $40$   & BH & --  \\
GM3 & $180$  & $30$ & $1.52$  & BH  &  $40$   &  $42$   & BH & --   \\

\end{tabular}
\caption{\label{grb}
Energy released $\Delta E$ in
the conversion to a hybrid or a quark star, for
various sets of model parameters. $M_{cr}$ is the gravitational mass
of the hadronic star at which the transition takes place, for fixed
values of the surface tension $\sigma$ and of the mean life-time
$\tau$ (here we have assumed $\tau=1$ year).}
\end{center}
\end{table}

By continuing mass accretion, the nucleation time can then be
reduced from values of the order of the age of the universe down to a
value of the order of days or years. We can therefore determine the
critical mass $M_{cr}$ of the metastable HS for which the nucleation
time corresponds to a fixed small value (1 year in Tab.~\ref{grb}).

In Table \ref{grb} we show the value of $M_{cr}$ for various sets of model
parameters. In particular GM3 refers to the hadronic equation of state \cite{glen2},
$B^{1/4}$ is the value of the MIT bag constant which is included in the equation of state
of quark matter and $\Delta_i$ are four different superconducting gaps of the Color-Flavor Locked (CFL) 
phase taken from Ref.~\cite{noiprd}.
In the conversion process from a metastable hadronic star into an
hybrid or a quark star a huge amount of energy $\Delta E$ is released. We
see in Table \ref{grb} that the formation of a CFL phase allows to obtain
values for $\Delta E$ which can be much larger
than the corresponding $\Delta E$ of the unpaired quark matter case
($\Delta=0$).

In the model we are presenting, the GRB is due to the cooling of the
justly formed hybrid or quark star via neutrino - antineutrino emission.  The
subsequent neutrino-antineutrino annihilation generates the GRB.
As shown in Ref.~\cite{Salmonson:1999es},
near the surface of a compact star, due to general relativity
effects, the efficiency of the neutrino-antineutrino annihilation is strongly
enhanced with respect to the Newtonian case.
In our scenario the duration of the prompt emission of the GRB is
therefore regulated by two mechanisms: 1) the time needed for the
conversion of the hadronic star into a hybrid or quark star, once a critical-size droplet is
formed and 2) the cooling time of the justly formed hybrid  or quark star.
Concerning the time needed for the conversion into quark matter of at least a
fraction of the star, it is possible to show
that the stellar conversion is a very fast process, having a duration
much shorter than 1s \cite{nuovo}. On the other hand, the neutrino trapping time,
which provides the cooling time of a compact star, is of the order of
$\sim 10$ s \cite{ignazio2}, and it gives the typical duration
of the GRB prompt emission in our model.\\

{\bf Temporal structure of GRBs}\\

The time structure of long GRBs is usually very complex \footnote{In
a sizable fraction of cases there is also evidence of a precursor activity before the GRB
prompt emission, with time delays up to 200 s \cite{lazzati}. In
Ref.~\cite{haensel} it has been speculated that the existence of precursors
can help in shedding light on the inner engine.}. In the light curves it 
is in fact possible to distinguish
several short pulses separated by time intervals lasting from
fractions of second to several ten of seconds.  From a statistical
analysis performed by Nakar and Piran \cite{nakar} it turns out that
the distribution of intervals between the peaks of the light curves is
well described by a log-normal distribution function up to delays of
roughly three seconds while for longer intervals a noticeable
deviation from the log-normal distribution is present. According to
Nakar and Piran, such a deviation occurs because a different mechanism
governs the high end tail of the distribution and it suggests that the
long quiescent times reflect periods in which the ``inner engine'' of
the GRB is not active, while the log-normal distribution correspond to
delays having a stochastic origin.

\begin{figure}[t]
    \begin{centering}
\hbox{\hskip-0.cm \includegraphics[scale=0.45]{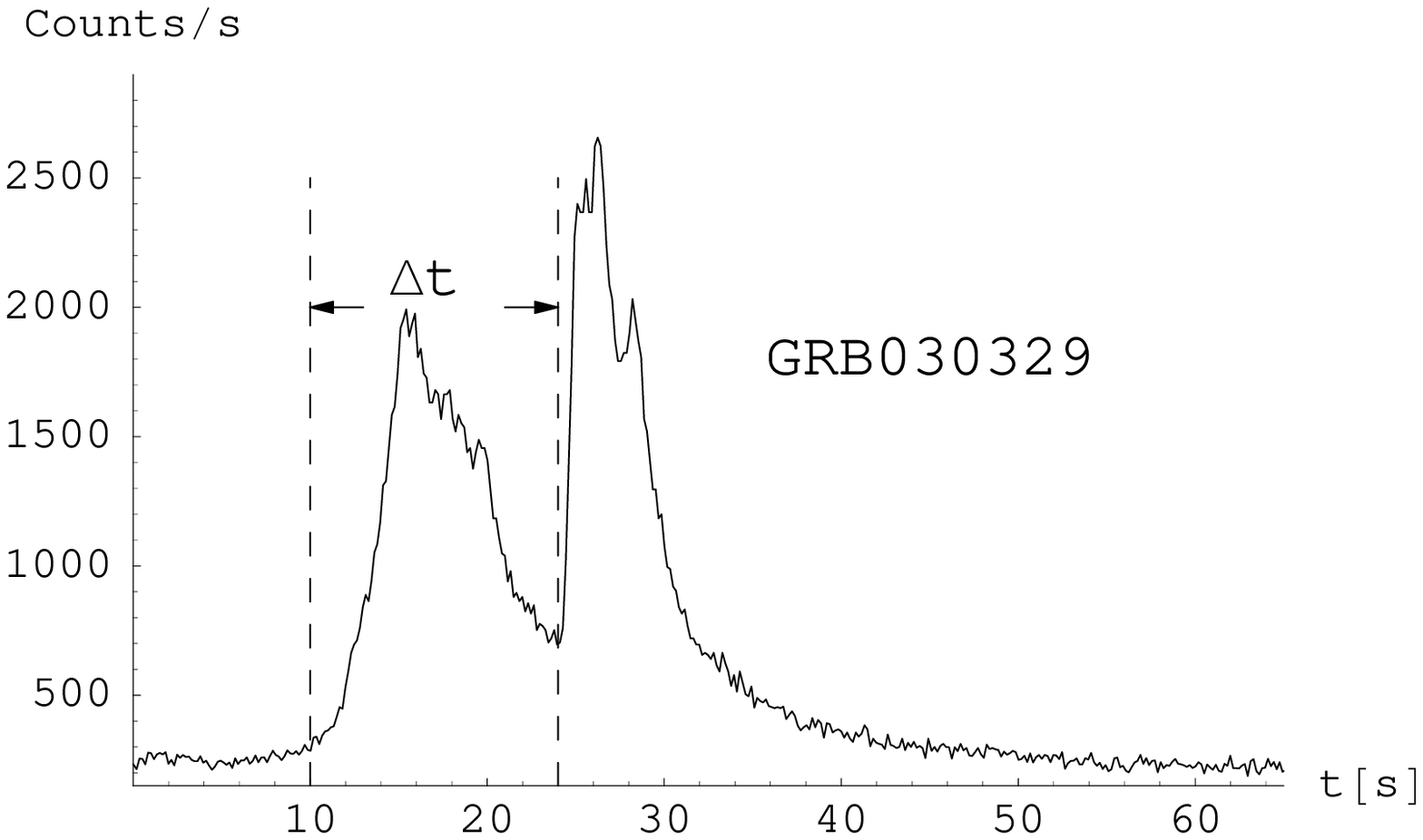} \hskip-0.cm \includegraphics[scale=0.45]{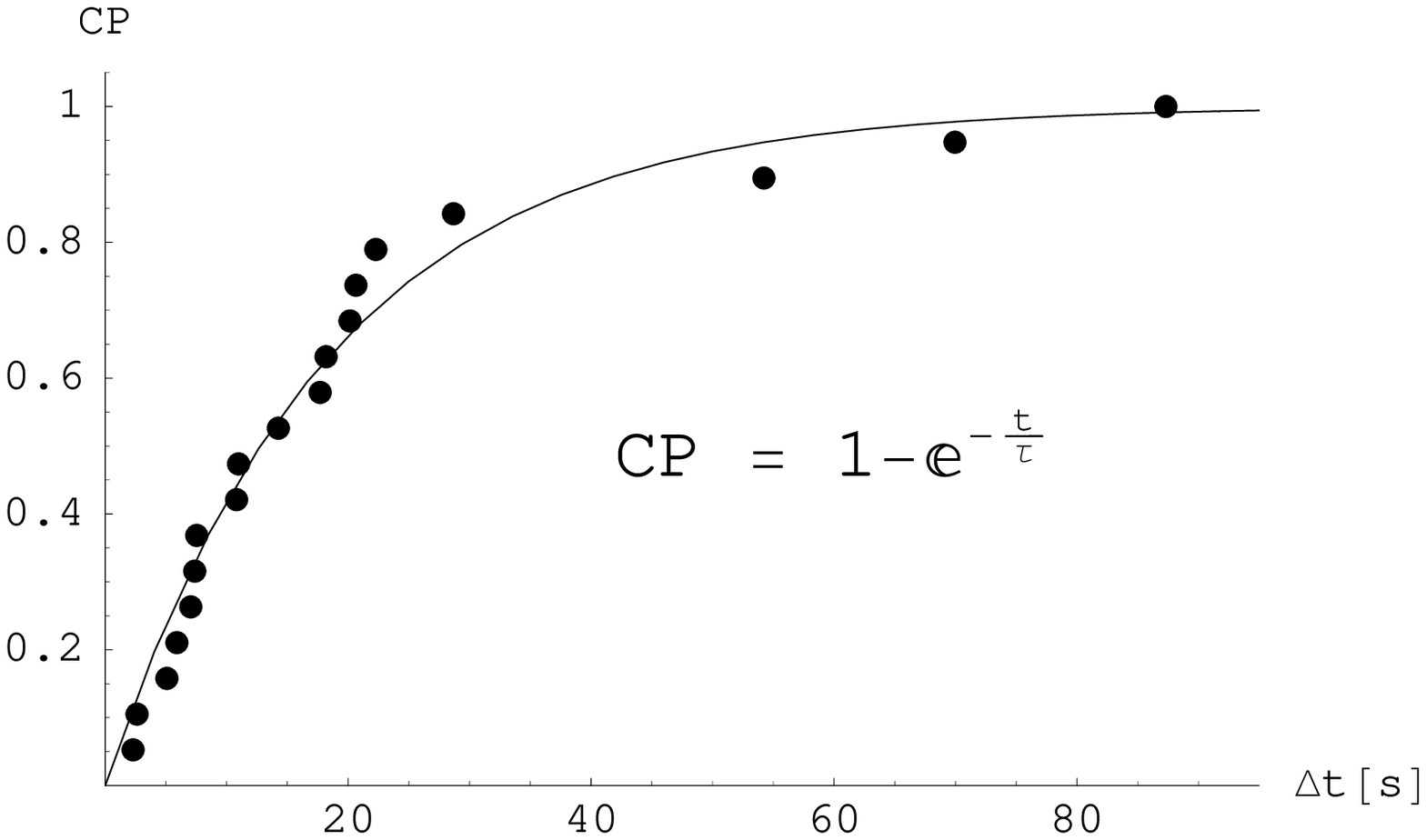} }
    \caption{Left panel: Light curve of GRB030329 from HETE catalogue. The two active periods of the GRB are
                         separated by a time interval $\Delta t \sim 14$ s.
Right panel: cumulative probability of $\Delta t$ from the GRBs of the HETE catalogue (dots) and the
exponential cumulative distribution (solid line).\label{hete} }
   \end{centering}
   \end{figure} 

In our model long periods of quiescence of the ``inner engine'' are
possible.  In the calculations presented in Table \ref{grb}, we
assumed a direct first order transition from
hadronic matter to the CFL phase. Actually, recent results on
the QCD phase diagram 
\cite{Alford:2004hz,Ruster:2005jc,Blaschke:2005uj,Lavagno}
suggest that the transition from hadronic to CFL phase can
proceed in two steps, first with a transition from
hadronic matter to a 2SC phase (or to unpaired quark matter,
depending on the model parameters) and then from 2SC to CFL.
In the scheme we are proposing, the first transition takes place
due to the increase of the baryonic density (due to mass accretion),
while the second transition is associated with the deleptonization
(and the cooling) of the newly formed star containing the 2SC phase.
These two transitions can both be first order \cite{lastshov}
and therefore the newly
formed hybrid or quark star containing 2SC quark matter can become
metastable and then decay into a star containing CFL phase with a
characteristic time delay which corresponds to the nucleation time of a drop of
CFL phase inside the 2SC phase.
In our model these time delays are
connected with the time intervals $\Delta t$ separating the active periods 
of long GRBs and therefore the delays $\Delta t$  should follow an exponential distribution. Active
periods correspond to emission periods during which the signal 
is about $4\sigma$ above the background \cite{NP2}.
It is important to notice that the first transition takes place when the central
density of the star reaches a ``critical'' density, whose numerical value depends
on the model parameters but, for a specific choice of the parameters value this density is 
determined to be in a very narrow range. From this viewpoint, the first
transition acts as a ``mass filter'' and therefore the second transition takes
place in a star which, in all bursts, has essentially always the same mass.

We tested our model performing a very simple statistical analysis on the 
small sample of GRBs detected by HETE. 
In the right panel of Fig.\ref{hete}, the cumulative distribution of $\Delta t$    
extracted from the analysis of the observations
is shown together with an exponential distribution $1-\exp(-t/\tau)$,
with $\tau \sim 20$ s. It is clear that the theoretical distribution
is compatible with the data (as it also results from the Kolmogorov-Smirnov test).
Unfortunately the HETE catalogue contains only a very small number of GRBs,
insufficient to reach a clear conclusion 
and therefore a statistical investigation based on the 
huge BATSE catalogue is now in progress.\\

It is a pleasure to thank D.~Lazzati and E.~Montanari for many 
useful discussions.

\end{document}